# Fault Characterization and Hardening of Combinational Standard Cells Using 3D-TCAD Simulations for Cyber-Physical Systems

Ali Zarei
*Department of Computer Engineering*
*Amirkabir University of Technology*
*(Tehran Polytechnic)*
Tehran, Iran
ali.zarei@aut.ac.ir

Amir M. Hajisadeghi
*Department of Computer Engineering*
*Amirkabir University of Technology*
*(Tehran Polytechnic)*
Tehran, Iran
hajisadeghi@aut.ac.ir

Hamid R. Zarandi
*Department of Computer Engineering*
*Amirkabir University of Technology*
*(Tehran Polytechnic)*
Tehran, Iran
h_zarandi@aut.ac.ir

***Abstract*— Cyber-physical systems (CPSs) are increasingly employed in applications with various levels of mission criticality, making the reliability of digital system components essential for maintaining service quality. On the other hand, advancements in technology nodes have heightened reliability concerns in these systems. This paper presents a method for characterizing and enhancing the fault tolerance of combinational standard cells using 3D-TCAD simulations. Through detailed simulations, we identify fault-sensitive regions in widely used standard cells under diverse scenarios that include variations in fault energy, particle angle, and the adjacency effect of identical and non-identical neighboring cells. Following this high-precision characterization, a hardened version of a universal logic NAND cell is proposed that mitigates its vulnerabilities. Simulation results demonstrate substantial improvements in resilience to particle-induced faults.**

## I. INTRODUCTION

Cyber-physical systems (CPSs), including industrial automation and control systems, rely on control devices that that interface directly with the physical world through sensors and actuators, managing real-world processes like production or power systems [1]. Ensuring the reliability of these systems is essential, as they often operate in mission-critical environments [2, 3]. Addressing soft errors—caused by transient faults due to particle strikes—is a primary concern in enhancing CPS resilience, as these errors can significantly impact system dependability [2, 3].

As transistor sizes decrease and operating voltages lower by advancing technology, circuits have become increasingly susceptible to soft errors from particle strikes [4]. Strikes at varying angles and energies can induce multiple effects, like multiple node corruptions and charge sharing, particularly in combinational circuits, where soft error rates increase linearly with frequency [4, 5]. In digital design, circuit layout factors like standard cell placement and adjacency play critical roles, as they can influence single-event transients (SETs) and multiple-event transients (METs), ultimately affecting reliability [6, 7].

While some device-level studies [8-10] examine particle strikes in standard cells, they often lack a comprehensive approach that considers all key parameters, such as particle energy, angle, and the effects of adjacent cell placement. For instance, [8] investigates single event upset cross-sections in NOR and NAND gates but omits particle angle effects. Study [9] explores fault-tolerant standard cell design without considering adjacency impacts, and [10] models particle diffusion yet excludes angular variations.

This paper addresses these gaps by performing a comprehensive 3D device-level simulation of particle strikes on combinational standard cells under varying energies, angles, and cell adjacency scenarios. The study evaluates critical characteristics like charge transfer and track length for both isolated and adjacent cells. Using 3D-TCAD simulations, we identify the fault susceptibility of INV, NAND, and NOR cells and their failure probabilities. Following this high-precision characterization, hardened versions of the universal NAND cells by hardening the more sensitive zones are proposed to mitigate these fault vulnerabilities. Simulation results demonstrate substantial improvements in fault immunity in resilience to particle-induced faults. The results provide design insights into optimizing CPS fault tolerance, particularly by suggesting standard cells with lower failure probabilities.

The paper is structured in the following manner. Section II is dedicated to previous works in sensitive areas determining and hardening. Section III presents the 3D-TCAD device-level simulations and metrics. Section IV proposes the design of a hardened cell and the fault resiliency results. Lastly, Section V presents the study's concluding remarks.

## II. RELATED WORK

Reliability analysis of combinational circuits under particle strikes has been extensively studied across multiple abstraction levels, with varying focus on fault localization, error propagation probability, and soft error mitigation techniques. Many gate-level methods prioritize rapid soft error propagation estimation. For example, [11] provides a fault propagation probability model considering logical, electrical, and timing masking factors, while [12] improves circuit reliability through graph-based restructuring and partitioning of circuits into sub-circuits. These gate-level approaches excel in analyzing soft error rates under both single and multiple event transients due to the ability to account for complete circuit masking effects.

At the transistor/layout level, multi-level techniques like those in [13-15] estimate and mitigate fault impacts by identifying the sensitive zones within combinational circuits. In [13], a multi-level approach achieves a 20% reduction in multiple event occurrences by identifying ion strike-affected cells and refining the layout to minimize sensitivity. Another technique [14] models multiple event transients in combinational cells with higher accuracy than gate-level simulations, showing that cell placement critically influences fault resilience and should be optimized during design.

Advancements in 3D-TCADs have expanded device-level analysis, allowing detailed study of physical effects like charge sharing in standard cells. In [8], different configurations (e.g., NOR-INV vs. OR, NAND-INV vs. AND) were analyzed, showing the impact of layout on single-event upset (SEU) cross-section. Follow-up works further enhance reliability through transistor stacking and gate sizing [16]. Sensitive area-based approaches, as in [9], use particle track simulations to quantify vulnerable regions and demonstrate that techniques like guard rings significantly mitigate single-event transients. The LBSEVEA method [10] combines device and SPICE-level simulations, analyzing factors like ambipolar diffusion and bipolar amplification to enhance cell vulnerability estimation.

While several studies examine single and multiple event effects on standard cells, others focus on factors like SET pulse width and pulse quenching [17, 18], using circuit-level simulation to characterize cell responses under transient conditions.

Based on our understanding, this work introduces the first combined device-level investigation of identical and non-identical adjacent standard cells, aiming to optimize fault tolerance without requiring space redundancy [6, 13]. We show that designers can achieve lower circuit failure probabilities by selecting cells with inherently lower failure rates and managing high-risk cell adjacencies. Our analysis includes extracting sensitive angles and areas for individual cells and adjacent cells, offering a comprehensive perspective for fault-tolerant design.

## III. 3D-TCAD Device-Level Simulation

This section provides an overview of the 3D-TCAD simulation environment and configuration, followed by an analysis of results highlighting the fault-prone areas in combinational standard cells. It emphasizes how particle strikes affect different cells and their adjacencies, providing data for subsequent hardening techniques.

### A. *Simulation Setup and Configurations*

Fig. 1 illustrates the schematic of the basic material structure of the transistor, which is evaluated within planar CMOS technology in 3D-TCAD. The simulations are conducted using GEANT4 [19], with all materials modeled according to 45*nm* FreePDK technology [20]. The drain and source length are set to $0.1\mu m$. Transistors are connected to a high-level voltage, with the supply voltage fixed at $1.0V$. The particle's linear energy transfer (LET) is considered constant throughout its travel within the material, with the four particle energy levels evaluated: 10, 50, 100, and 150*MeV* [21]. The particle is modeled with a length of $5\mu m$ and a radius of $0.05\mu m$ [22, 23].

To account for different transistor placements, the trail model is applied in obtaining simulation results [24]. Particle strikes are simulated in five key regions: 1) Source, 2) Lightly Doped Source, 3) Gate, 4) Lightly Doped Drain, and 5) Drain. Each region undergoes angular particle hits to identify the most critical impact angles and areas. The particle's incident angle ranges from 10° to 90° in a 2D front-face view.

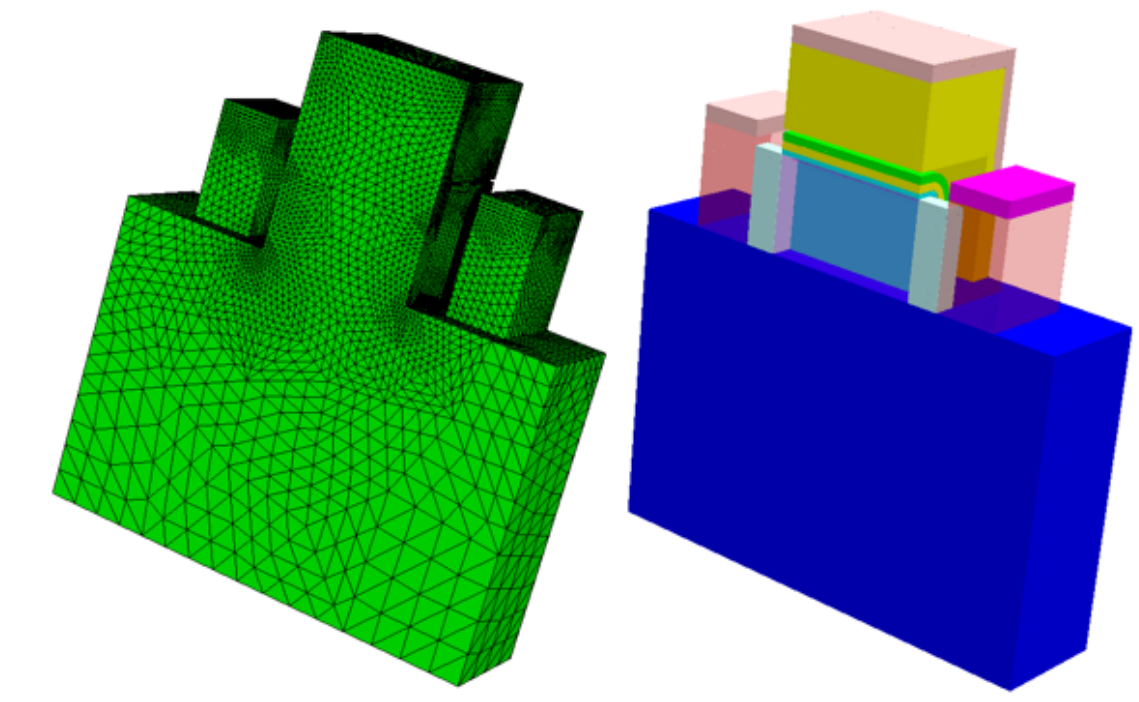

Fig. 1. The schematic of the basic material structure of the transistor in 3D-TCAD simulations.

The following physical models are implemented in this study [22, 23]:

- Fermi-Dirac distribution for statistical analysis.
- Band-gap narrowing.
- Recombination effects in Auger and Shockley-Read-Hall (SRH) processes.
- Mobility dependencies on temperature, doping levels, electric fields, and carrier-carrier scattering.
- Heavy ion tracking and incident modeling, utilizing a Gaussian radial profile.
- Carrier transport is modeled with the Hydrodynamic model.

Finally, due to limited prior investigation on ion impacts specific to standard cells (particularly considering variations in hit position and angle) a magnifier was set up on standard cells for closer analysis.

While the use of 45nm FreePDK technology in our simulations ensures compatibility with open-source models and GEANT4 integration, the behavior of soft errors can differ significantly in advanced technology nodes such as 7nm or 5nm. These nodes often employ FinFET structures and exhibit stronger electric fields and reduced node capacitance, potentially amplifying charge sharing effects and increasing sensitivity to particle strikes. Adapting the current methodology to such nodes is an important direction for future work, particularly considering emerging applications in low-power and high-performance CPS environments.

### B. *Device-level Investigation Results*

Figs. 2 (a) through (e) present the collected charge results from particle strikes at varying angles across the five distinct zones of the transistor. From these results, two key insights are obtained. First, angles approaching the terrain demonstrate increased energy transfer, resulting in greater vulnerability to charge sharing between nodes. Across all zones, the 10° angle near the terrain consistently shows a deeper impact than other angles. Second, in terms of susceptible zones, while the drain is commonly identified as the primary vulnerable zone that accumulates charge and can flip the output, our results also indicate that the source can be susceptible.

Given the significant impact of layout design on fault characterization—particularly charge sharing between closely placed cells—our study further investigates different combinations of identical and non-identical adjacent cells, analyzing their sensitive radius. The sensitive radius is defined as the radial distance within which neighboring cells can be affected. Note that we have chosen the standard layout of the 45*nm* NanGate open cell library [24] to obtain adjacent cell simulation results and new layout designs could affect the cell sensitivity to particle strike.

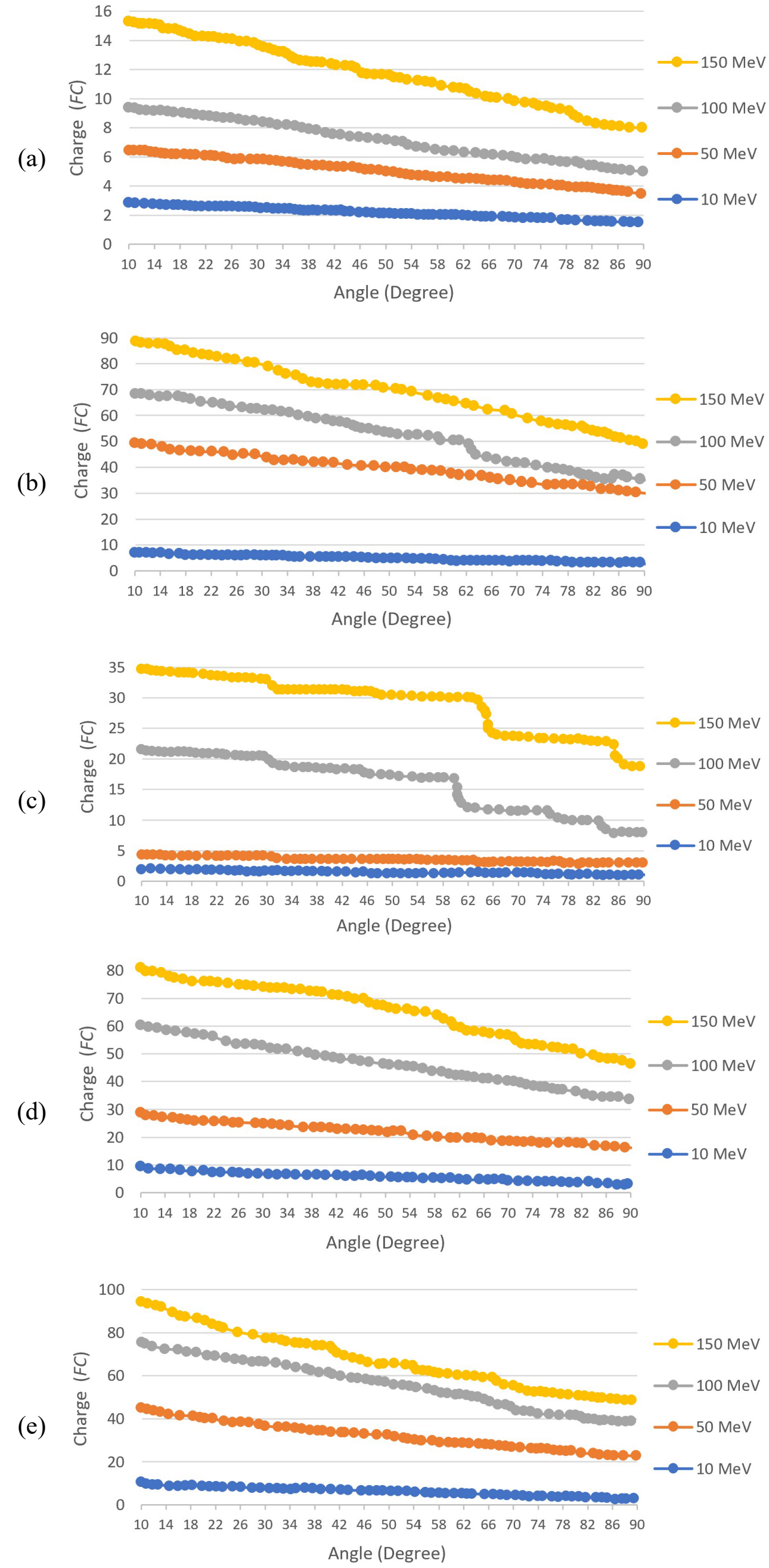

Fig. 2. Charge collected in the different areas of transistor; a) source, b) lightly doped source, c) gate, d) drain, e) lightly doped drain.

Initially, we examine particle strikes on identical standard cells which means each cell is surrounded by identical counterparts. The sensitive radius results for an INV cell surrounded by INV cells, under particle strikes with energies of 50*MeV* and 100*MeV* in sensitive NMOS and PMOS transistor zones, are shown in Fig. 3. Also, results for NAND and NOR cells in identical surroundings are displayed in Figs. 4 and 5, respectively.

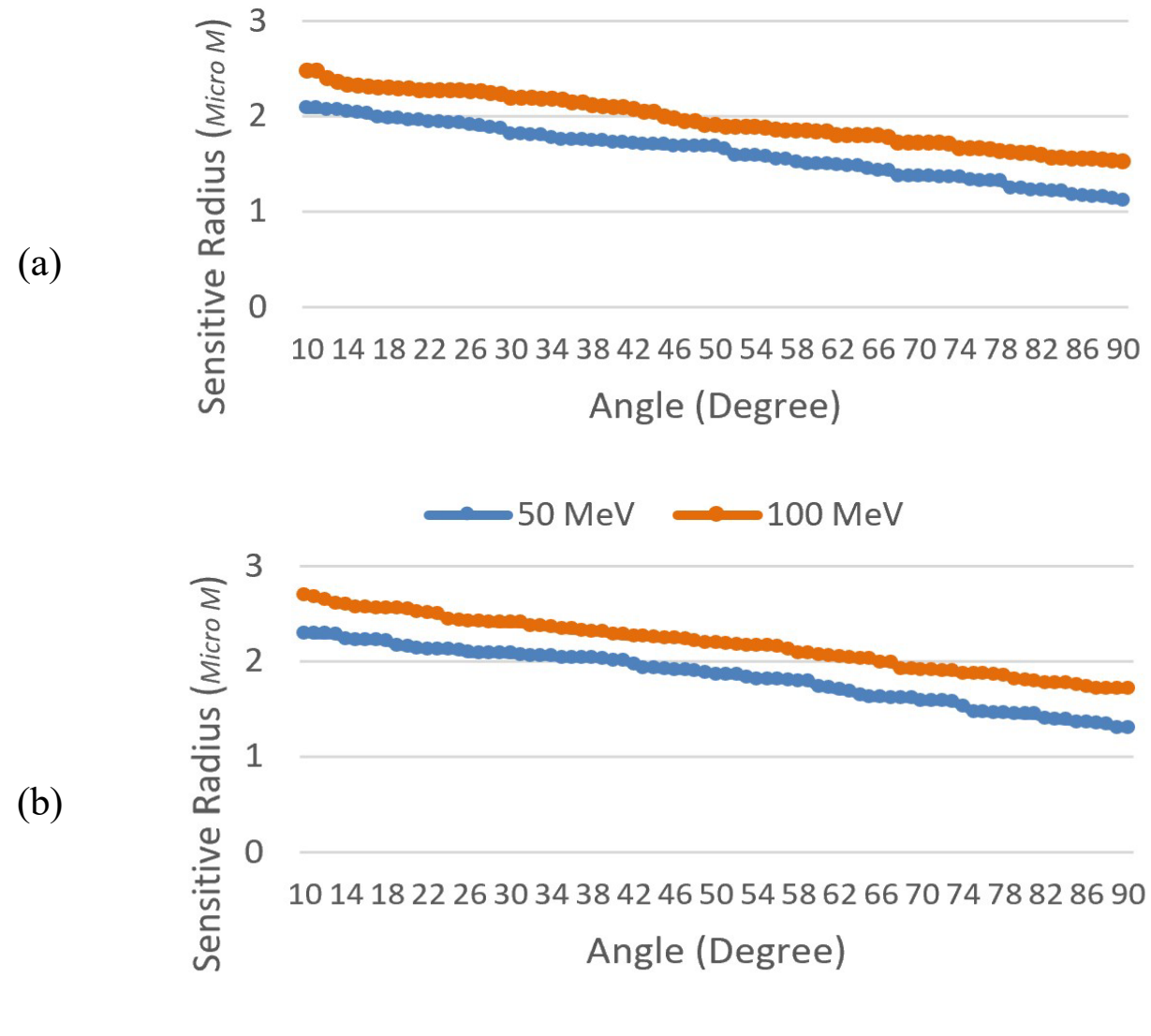

Fig. 3. Sensitive radius of INV cell; a) NMOS, b) PMOS transistors.

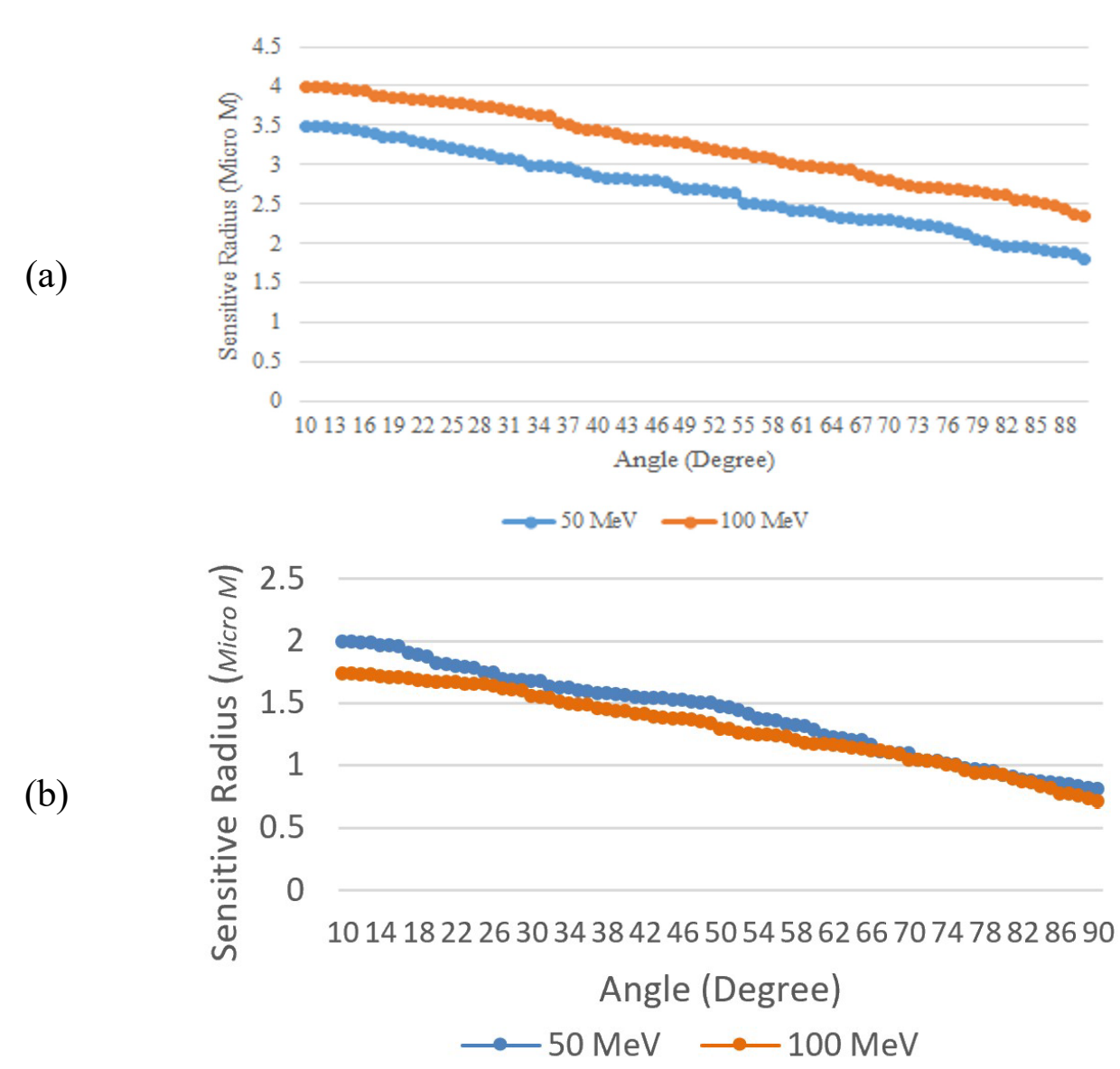

Fig. 4. Sensitive radius of NAND cell; a) NMOS, b) PMOS transistors.

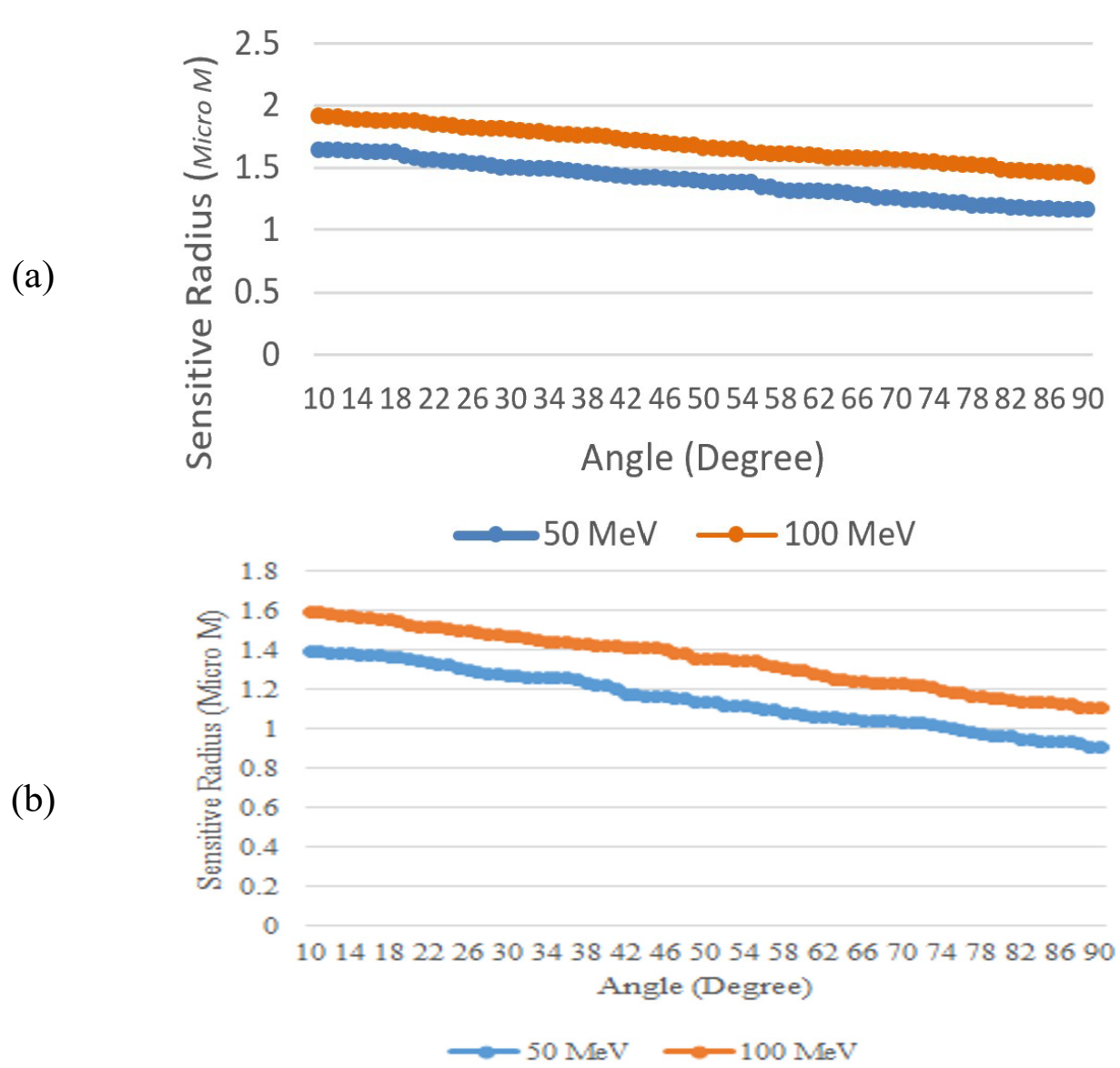

Fig. 5. Sensitive radius of NOR cell; a) NMOS, b) PMOS transistors.

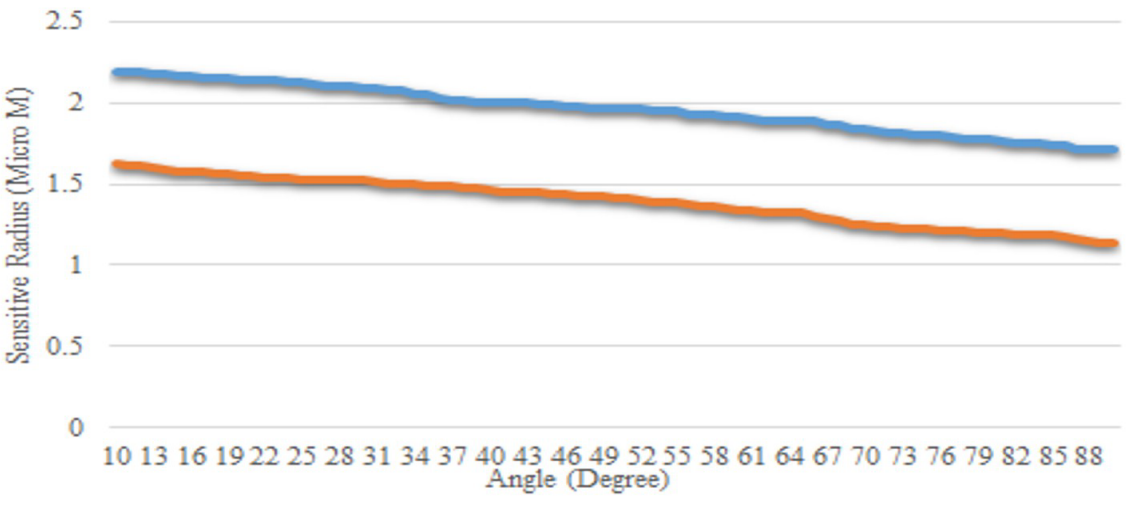

Fig. 6. Sensitive Radius of INV Cell among INV and NOR cells.

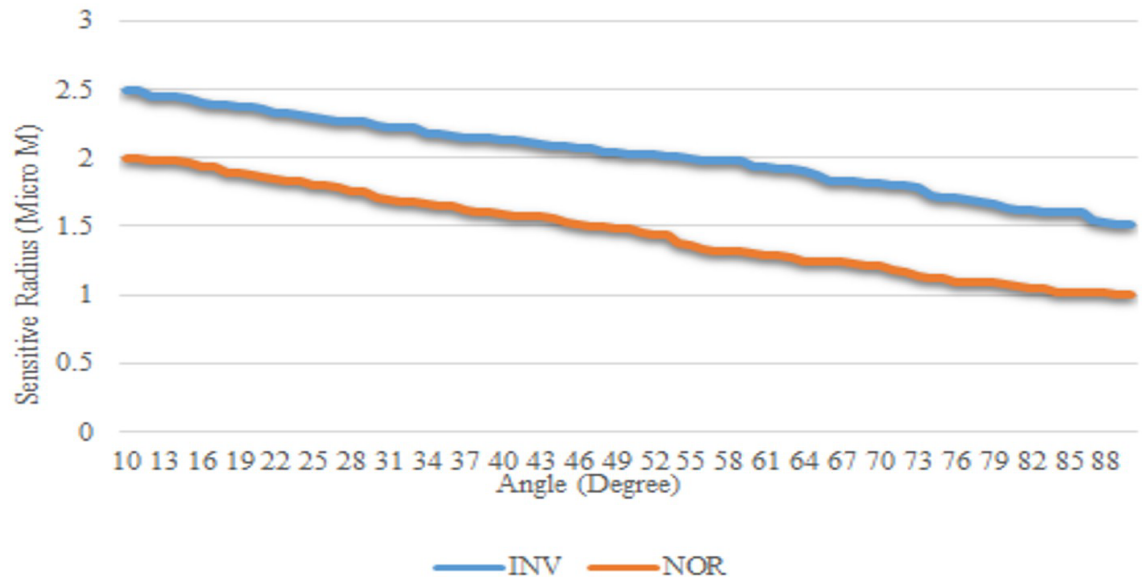

Fig. 7. Sensitive Radius of NAND Cell among INV and NOR cells.

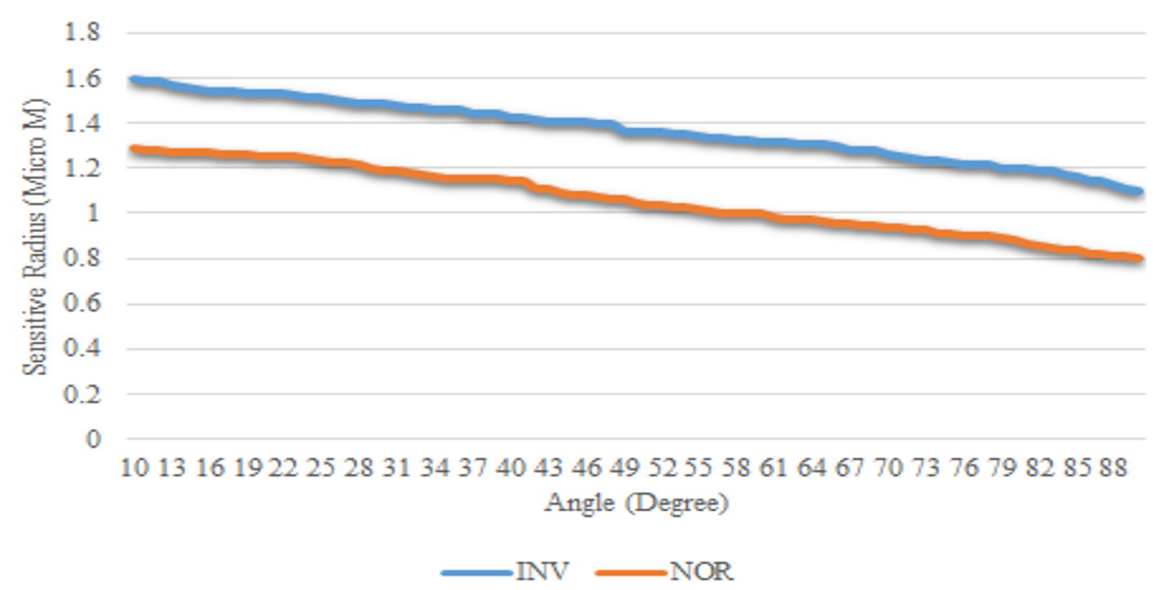

Fig. 8. Sensitive Radius of NOR Cell among INV and NOR cells

These simulations reveal that, while all three standard cells are susceptible to particle strikes, the NAND cell exhibits the largest sensitive radius, making it the most prone to energy collection and sharing. The INV cell follows in vulnerability, while the NOR cell is the least affected. The findings affirm that fault vulnerability is not only contingent on sensitive zones but also significantly influenced by the cell structure, due to variability in the distance between sensitive zones.

Since in real-world digital systems frequently cells surrounded by non-identical neighbors to enhance reliability, we also investigated non-identical adjacent cell configurations. Simulations were conducted on various cell combinations, with an INV cell surrounded by NAND and NOR cells and vice versa. For each configuration, the average sensitive radius under particle energies of $50MeV$ and $100MeV$ at different angles was calculated for INV, NAND, and NOR cells as target cells, shown in Figs. 6, 7, and 8.

The results show that when the INV cell is centered among NAND and NOR cells, both neighboring cells are susceptible, with the NAND cell being more vulnerable due to its larger sensitive radius. Also, when the NAND cell is centered, INV cells surrounding it exhibit higher vulnerability than NOR cells. Lastly, with the NOR cell at the center, the surrounding NAND cells demonstrate greater susceptibility compared to INV cells. The sensitive radius for NOR-centered configurations is lower than for the other cases, suggesting that NOR cells possess inherent resilience to particle strikes, likely due to structural attributes that attenuate energy transfer.

The greater vulnerability observed in NAND cells is primarily due to their structural layout, where the series connection of NMOS transistors and the proximity of sensitive nodes increase the likelihood of fault propagation. Additionally, the energy collected in these zones during a particle strike can more easily impact the output logic, especially under certain input combinations. These structural characteristics, along with reduced node separation, make the NAND cell more susceptible compared to NOR and INV cells.

In summary, combining layout information with sensitive radius analysis effectively highlights the susceptibility of cells in both identical and non-identical adjacent configurations. Fig. 9 provides a summary of configurations, illustrating the probability of surrounding cells being affected when the center cell is struck by particle.

Despite the high accuracy of GEANT4 and 3D-TCAD in modeling physical-level phenomena, some limitations should be noted. The assumption of constant LET may not capture variations in real-world radiation environments. Moreover, effects such as process variation, interconnect parasitic, and temperature fluctuations are not modeled in this study. These factors could affect the actual fault susceptibility in fabricated circuits. Future work may incorporate statistical modeling and post-layout simulation to further validate the results.

## IV. Design Hardened Cell

Previous approaches to hardening standard cells against particle strikes often rely on spatial redundancies [6, 13], which require precise placement data as the final cell positions are determined post-placement. As a result, changes to placement can affect the efficacy of these methods. In contrast, this study introduces a hardened version of standard cells designed through physical 3D-TCAD simulation to specifically target and harden vulnerability zones.

Through 3D-TCAD simulations, we identified sensitive nodes within standard cells, with NAND cells exhibiting the highest vulnerability among those analyzed. Given its fundamental role as a universal logic gate, the NAND cell was prioritized for hardening. Our analysis showed that the NAND cell has two primary zones of vulnerability, as illustrated in Fig. 10 (a). Particle strikes in these zones significantly increase the chance of faults affecting the output, depending on the input state.

The hardened NAND cell design, shown in Fig. 10 (b), includes three additional transistors to redirect and neutralize charge in fault scenarios under various input conditions. In this design, if a particle strike occurs in one of the vulnerable areas, the collected charge is dissipated through alternative paths to the ground or source. This mechanism ensures that, based on the input state and prior logic level of the affected node, the faulted node is either clamped to GND or VDD-$V_t$, effectively preventing charge from impacting the output node.

For clarification, consider two scenarios with particle strikes on the output node. First, when both inputs are “11,” any hit affecting the PMOS transistors could potentially propagate a fault to the output. To address this, an additional PMOS transistor has been introduced to the drains of the PMOS transistors, which operates similarly for inputs "01" or "10." In the second scenario, with both inputs at “00,” the pull-up network remains active (VDD), while the pull-down is inactive. Here, if the strike affects a PMOS transistor, the output voltage may temporarily exceed VDD; if it affects an NMOS transistor, the drop will remain within the noise margin, preserving correct logic functionality. Even with high LET particle energy of $150MeV$, simulations show only a minimal voltage decrease from $1V$ (VDD) to $0.77V$, which is maintained as a logical high in the hardened circuit.

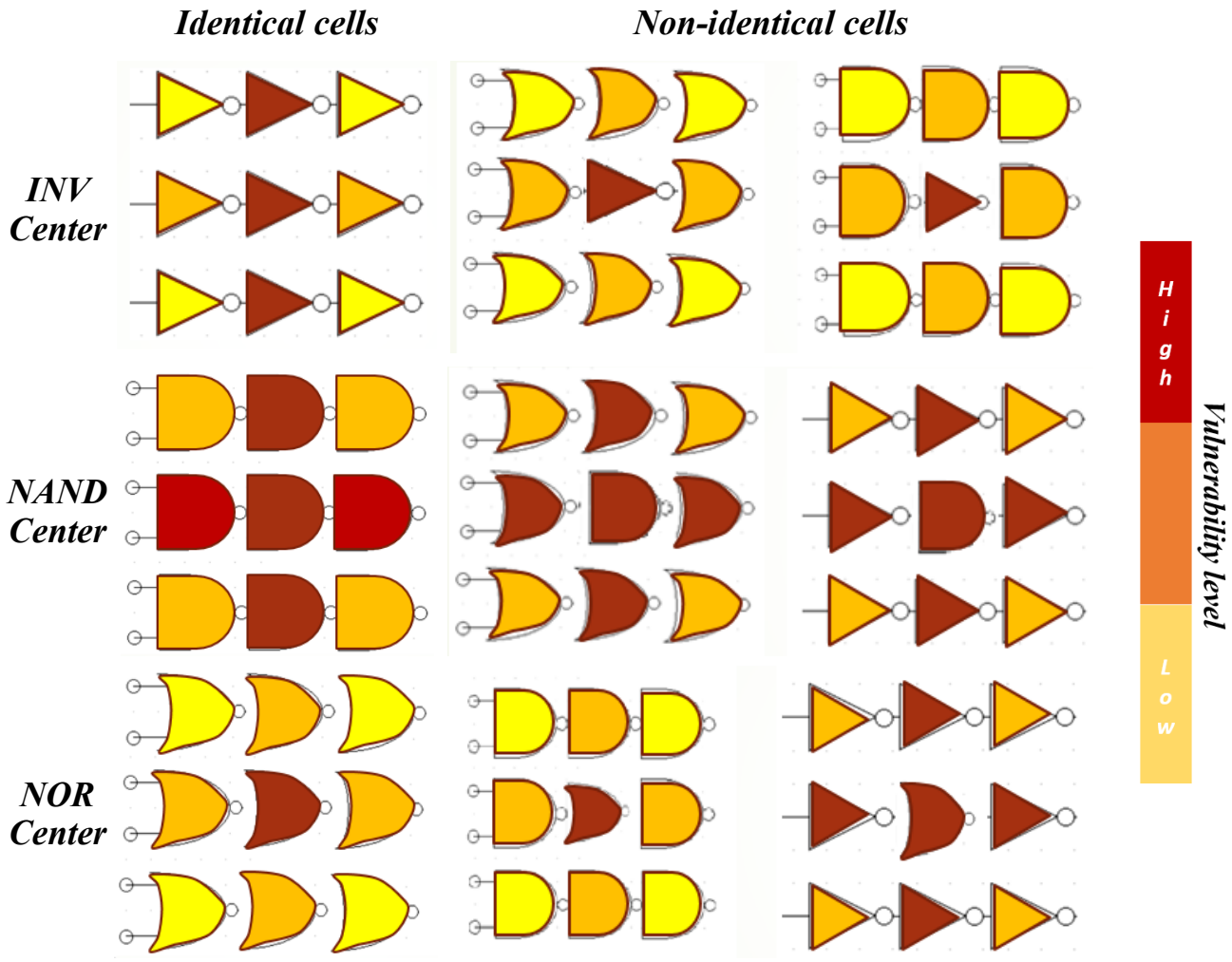

Fig. 9. Vulnerability level of INV, NAND, and NOR cells among other cells when particle strikes to the center cell.

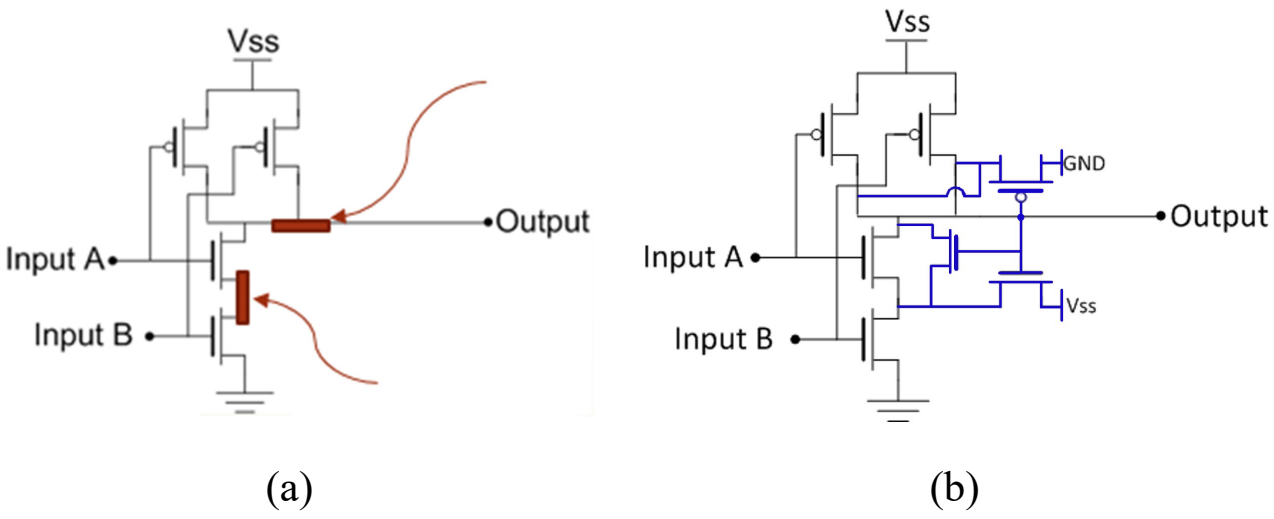

Fig. 10. a) Sensitive areas, and b) the hardened design of NAND cell.

To quantify the hardened NAND cell's fault tolerance, we conducted extensive SPICE simulations, utilizing the Piecewise Linear (PWL) current model to emulate particle strikes at various energy levels. Fig. 11 compares the mean fault propagation probability between the baseline and hardened NAND cells. For particle energies under $50MeV$, the hardened NAND cell's fault propagation probability remains below 5%, and for higher energies, it remains under 11%. Despite the overhead of three additional transistors, the hardened design significantly reduces fault propagation probabilities across varying particle energies.

On average, The hardened NAND cell design achieves 92.6% reductions in fault propagation probability over the baseline NAND cell, offering a robust solution for applications requiring high reliability, such as cyber-physical systems with diverse mission-critical requirements.

The proposed hardened NAND cell introduces three additional transistors, resulting in a ~37.5% increase in the total transistor count compared to a conventional 6-transistor NAND gate. Preliminary SPICE-level simulations show a marginal increase of ~6% in propagation delay, with negligible impact on power consumption under typical switching activity. The cell maintains compatibility with 45nm standard design rules and requires no additional area if integrated within existing placement constraints. These results highlight a favorable trade-off between reliability and performance.

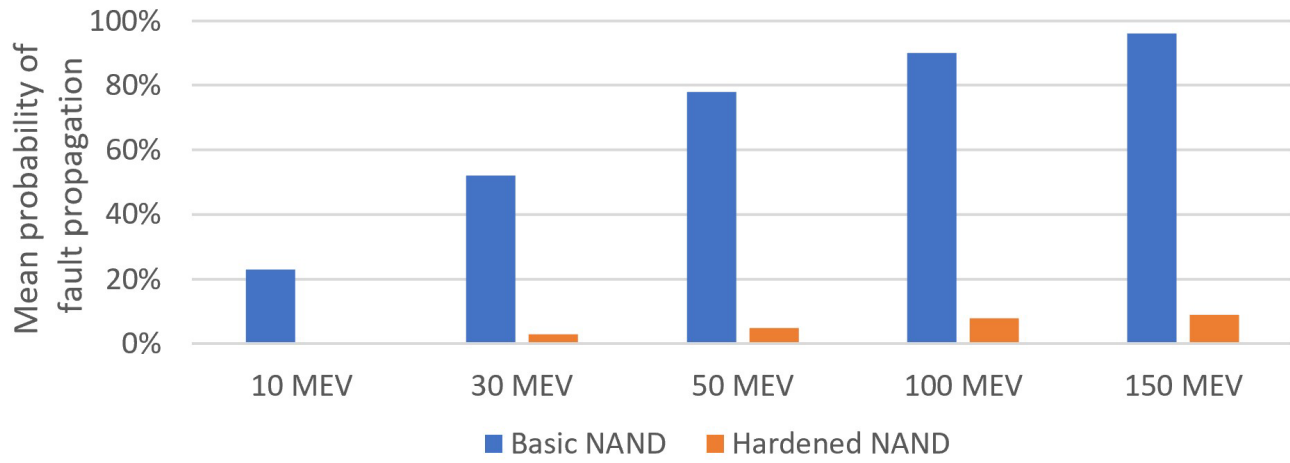

Fig. 11. Mean probability of fault propagation in baseline and hardened NAND cells.

It is worth mentioning that traditional radiation-hardening techniques such as Triple Modular Redundancy (TMR), gate sizing, transistor stacking, and guard rings have been widely explored for improving fault resilience in standard cells [6], [13], [16]. However, most of these methods either rely on spatial redundancy or require specific layout constraints. In contrast, our proposed design targets the physical vulnerability zones directly within the cell using insights from 3D-TCAD simulations, eliminating the need for post-layout adjustments. This leads to a layout-compatible and technology-independent solution with lower area overhead compared to redundancy-based methods.

## V. Conclusion

In this work, we presented a comprehensive fault characterization and hardening approach for combinational standard cells using 3D-TCAD device-level simulations. Our study evaluated fault propagation across various angles, energies, and strike zones within standard cells, identifying the most vulnerable areas and assessing fault sensitivity in both identical and non-identical cell arrangements. Based on these insights, we proposed a hardened NAND cell design, adding targeted transistors to neutralize charge effects and reduce fault propagation probability under different particle energies. SPICE simulations validated that the hardened NAND cell significantly minimizes fault impact with acceptable overhead, making it well-suited for high-reliability applications. The future direction for this work involves developing hardened versions of additional standard cells.